\documentclass[useAMS,usenatbib]{mn2e}
\usepackage{epsf,graphicx,rotating,url}

\title[Coronal line variability in NGC 4151]{Variability of the coronal line region in NGC~4151}

\author[H. Landt et al.]{Hermine Landt$^1$\thanks{E-mail:
    hermine.landt@durham.ac.uk}\thanks{Visiting Astronomer at the
    Infrared Telescope Facility, which is operated by the University
    of Hawaii under Cooperative Agreement no. NNX-08AE38A with the
    National Aeronautics and Sp ace Administration, Science Mission
    Directorate, Planetary Astronomy Program.}, Martin J. Ward$^1$\footnotemark[2], Katrien C. Steenbrugge$^{2,3}$ and Gary J. Ferland$^{4,5}$ \\
$^1$Department of Physics, Durham University, South Road, Durham DH1 3LE, UK \\
$^2$Instituto de Astronom\'ia, Universidad Cat\'olica del Norte, Avenida Angamos 0610, 1270709 Antofagasta, Chile \\
$^3$Department of Physics, University of Oxford, Keble Road, Oxford OX1 3RH, UK \\
$^4$School of Mathematics and Physics, Queen's University of Belfast, Belfast BT7 1NN, Northern Ireland, UK \\
$^5$Department of Physics and Astronomy, University of Kentucky, Lexington, KY 40506, USA}

\begin{document}

\def\la{\mathrel{\hbox{\rlap{\hbox{\lower4pt\hbox{$\sim$}}}\hbox{$<$}}}}
\def\ga{\mathrel{\hbox{\rlap{\hbox{\lower4pt\hbox{$\sim$}}}\hbox{$>$}}}}

\font\sevenrm=cmr7

\def\SII{[S~{\sevenrm II}]}
\def\SIII{[S~{\sevenrm III}]}
\def\SVIII{[S~{\sevenrm VIII}]}
\def\SIX{[S~{\sevenrm IX}]}
\def\SXI{[S~{\sevenrm XI}]}
\def\SXII{[S~{\sevenrm XII}]}

\def\SiII{[Si~{\sevenrm II}]}
\def\SiIII{[Si~{\sevenrm III}]}
\def\SiVI{[Si~{\sevenrm VI}]}
\def\SiVII{[Si~{\sevenrm VII}]}
\def\SiVIII{[Si~{\sevenrm VIII}]}
\def\SiIX{[Si~{\sevenrm IX}]}
\def\SiX{[Si~{\sevenrm X}]}
\def\SiXI{[Si~{\sevenrm XI}]}

\def\OI{[O~{\sevenrm I}]}
\def\OII{[O~{\sevenrm II}]}
\def\OIII{[O~{\sevenrm III}]}

\def\OVII{O~{\sevenrm VII}}
\def\OVIII{O~{\sevenrm VIII}}
\def\OIX{O~{\sevenrm IX}}

\def\NeIX{Ne~{\sevenrm IX}}

\def\NVI{N~{\sevenrm VI}}
\def\NVII{N~{\sevenrm VII}}

\def\CIV{C~{\sevenrm IV}}
\def\CVI{C~{\sevenrm VI}}
\def\CVII{C~{\sevenrm VII}}

\def\FeII{[Fe~{\sevenrm II}]}
\def\FeVI{[Fe~{\sevenrm VI}]}
\def\FeVII{[Fe~{\sevenrm VII}]}
\def\FeX{[Fe~{\sevenrm X}]}

\def\FeVIIp{Fe~{\sevenrm VII}}

\def\cloudy{{\sevenrm CLOUDY}}

\date{Accepted ~~. Received ~~; in original form ~~}

\pagerange{\pageref{firstpage}--\pageref{lastpage}} \pubyear{2015}

\maketitle

\label{firstpage}

\begin{abstract}

We present the first extensive study of the coronal line variability
in an active galaxy. Our data set for the nearby source NGC~4151
consists of six epochs of quasi-simultaneous optical and near-infrared
spectroscopy spanning a period of about eight years and five epochs of
X-ray spectroscopy overlapping in time with it. None of the coronal
lines showed the variability behaviour observed for the broad emission
lines and hot dust emission. In general, the coronal lines varied only
weakly, if at all. Using the optical \FeVII~and X-ray \OVII~emission
lines we estimate that the coronal line gas has a relatively low
density of $n_e \sim 10^3$~cm$^{-3}$ and a relatively high ionisation
parameter of $\log U \sim 1$. The resultant distance of the coronal
line gas from the ionising source is about two light years, which puts
this region well beyond the hot inner face of the obscuring dusty
torus. The high ionisation parameter implies that the coronal line
region is an independent entity rather than part of a continuous gas
distribution connecting the broad and narrow emission line regions. We
present tentative evidence for the X-ray heated wind scenario of Pier
\& Voit. We find that the increased ionising radiation that heats the
dusty torus also increases the cooling efficiency of the coronal line
gas, most likely due to a stronger adiabatic expansion.

\end{abstract}


\section{Introduction}

In addition to the broad and narrow emission lines, the spectra of
active galactic nuclei (AGN) display high-ionisation emission lines,
the so-called coronal lines, which require energies $\ga 100$~eV to be
excited. The coronal line region is believed to lie at distances from
the central ionising source intermediate between those of the broad
(BELR) and narrow emission line region (NELR) and to possibly coincide
with the hot inner face of the circumnuclear, obscuring dusty torus
\citep[as first suggested by][]{Pier95}. Support for this assumption
comes from the fact that the coronal lines have higher critical
densities for collisional deexcitation than the low-ionisation narrow
emission lines ($n_e \sim 10^7 - 10^{10}$~cm$^{-3}$), their profiles
tend to have full widths at half maxima (FWHM) intermediate between
those of the broad and narrow emission lines \citep[FWHM$\sim
  500-1500$~km~s$^{-1}$; e.g.,][]{Pen84, App88, Erk97, Rod02c, Rod11}
and the emission from this region is often extended but much less so
than that from the low-ionisation NELR \citep[on scales of $\sim
  80-150$~pc; e.g.,][]{Prieto05, Mueller06, Mueller11, Maz13}. Coronal
lines are observed with smilar frequency in both types of AGN
\citep{Ost77, Koski78}, but type 1 AGN have stronger coronal line
emission relative to their low-ionisation narrow lines than type 2 AGN
\citep{Mur98}. Therefore, it is likely that the coronal line region
has two components, one compact and one spatially extended, with only
the latter remaining unobscured by the dusty torus in type 2 AGN.

The high ionisation potentials of the coronal lines can be produced
either in a hot, collisionally ionised plasma, as is the case for the
solar corona from which these lines have their name, or in a gas
photoionised by the hard continuum of the AGN. In the first case, the
electron temperatures would be of the order of $T_e \approx 10^6$~K
and in the second case much lower ($T_e \sim 10^4-10^5$~K). Currently,
photoionisation is favoured, since for most AGN the observed flux
ratios between different coronal lines can be reproduced within a
factor of $\sim 2-3$ by these models \citep{Oliva94, Ferg97}, whereas
the temperature of the hot plasma would need to be fine-tuned within a
very narrow range \citep{Oliva94}. In any case, the coronal line
region is most likely dust free, since strong emission from refractory
elements such as iron, silicium and calcium are observed, which would
be severly reduced in a dusty environment.

The coronal lines are often blueshifted relative to the low-ionisation
narrow lines \citep[e.g.,][]{Pen84, Erk97, Rod02c}, which indicates
that the coronal line gas is in outflow. However, as \citet{Mull09}
concluded for the source Ark~564, the potential coronal line emitting
clouds must have undergone acceleration to the observed velocities
prior to these lines being produced, i.e. they have reached their
terminal velocity against the opposing drag and gravitational
forces. Then, given also the similar estimated physical conditions and
location, it has been proposed that the partly ionised gas that
produces the \OVII~ and \OVIII~absorption lines and edges seen in the
soft X-ray spectra of many AGN, i.e., the so-called 'warm absorber',
produces also the coronal lines in its (colder) outer regions
\citep{Netzer93b, Erk97, Por99}. In this wind model for the coronal
line region, a considerable contribution from shock ionisation is
expected.

The most stringent constraints on the properties of the coronal line
emitting region could come from variabililty studies, in particular if
the variability of several coronal lines can be compared with each
other and with that of other AGN components such as the BELR and the
X-ray continuum. However, mainly due to the weakness of these emission
lines and also lack of data, very few studies of this kind have been
attempted so far. \citet{Vei88} did the only systematic study of the
coronal line variability. In his sample of $\sim 20$ AGN he found firm
evidence that both the \FeVII~$\lambda 6087$ and \FeX~$\lambda 6375$
emission lines varied (during a period of a few years) for only one
source (NGC~5548) and tentative evidence for another seven sources
(including NGC~4151). Then, within a general optical variability
campaign on the source Mrk~110 lasting for half a year,
\citet{Kolla01} reported strong \FeX~variations. More recently,
follow-up optical spectroscopy of a handful of objects with unusually
prominent coronal lines selected from the Sloan Digital Sky Survey
(SDSS) showed that in half of them the coronal lines strongly faded
(by factors of $\sim 2-10$), making these sources candidates for
stellar tidal disruption events \citep{Kom09, Yang13}.

In this paper, we present the first extensive study of the coronal
line variability in an AGN. Our data set for the nearby, well-known
source NGC~4151 ($z=0.0033$) is unprecedented in that it includes six
epochs of quasi-simultaneous optical and near-IR spectroscopy spanning
a period of $\sim 8$~years and five epochs of X-ray spectroscopy
overlapping in time with it. Furthermore, the observations in each
wavelength were performed with the same telescope and set-up. This
paper is organised as follows. In Section 2, we present the data and
measurements. In Section 3, we discuss the observed variability
behaviour of the near-IR, optical and X-ray coronal lines, for which
we seek an interpretation in the context of the location of and
excitation mechanism for the coronal line emission region in Section
4. Finally, in Section 5, we summarise our main results and present
our conclusions. Throughout this paper we have assumed cosmological
parameters $H_0 = 70$ km s$^{-1}$ Mpc$^{-1}$, $\Omega_{\rm M}=0.3$,
and $\Omega_{\Lambda}=0.7$.

\section{The data and measurements}

\subsection{Near-IR and optical spectroscopy}

\begin{table*}
\caption{\label{irlines} 
Near-IR emission line fluxes and ratios}
\begin{tabular}{lccccccccc}
\hline
Observation & \SIII~$\lambda$9531 & \SIII~$\lambda$9069 & \SIII/ & \SVIII~$\lambda$9911 & \SIII/ & \SIX~1.252$\mu$m & \SIII/ \\
Date & (erg/s/cm$^2$) & (erg/s/cm$^2$) & \SIII & (erg/s/cm$^2$) & \SVIII & (erg/s/cm$^2$) & \SIX \\
\hline
2002 Apr 23 & (7.61$\pm$0.09)e$-$13 & (2.66$\pm$0.02)e$-$13 & 2.86$\pm$0.04 & (2.49$\pm$0.21)e$-$14 & 30.6$\pm$2.6 & (2.44$\pm$0.12)e$-$14 & 31.2$\pm$1.6 \\
2004 May 23 & (1.06$\pm$0.02)e$-$12 & (3.89$\pm$0.08)e$-$13 & 2.73$\pm$0.08 & (5.35$\pm$0.60)e$-$14 & 19.8$\pm$2.3 & (2.64$\pm$0.28)e$-$14 & 40.2$\pm$4.3 \\
2006 Jan  8 & (1.17$\pm$0.01)e$-$12 & (4.40$\pm$0.06)e$-$13 & 2.66$\pm$0.04 & (3.81$\pm$0.26)e$-$14 & 30.7$\pm$2.1 & (3.41$\pm$0.14)e$-$14 & 34.3$\pm$1.4 \\
2006 Jun 12 & (9.97$\pm$0.10)e$-$13 & (3.85$\pm$0.07)e$-$13 & 2.59$\pm$0.05 & (3.38$\pm$0.19)e$-$14 & 29.5$\pm$1.7 & (3.54$\pm$0.12)e$-$14 & 28.2$\pm$1.0 \\
2007 Jan 24 & (8.17$\pm$0.14)e$-$13 & (2.79$\pm$0.06)e$-$13 & 2.93$\pm$0.08 & (3.09$\pm$0.38)e$-$14 & 26.4$\pm$3.3 & (3.01$\pm$0.29)e$-$14 & 27.1$\pm$2.6 \\
2010 Feb 27 & (8.89$\pm$0.15)e$-$13 & (2.98$\pm$0.06)e$-$13 & 2.98$\pm$0.08 & (4.00$\pm$0.36)e$-$14 & 22.2$\pm$2.0 & (3.31$\pm$0.19)e$-$14 & 26.9$\pm$1.6 \\
\hline
Observation & \SIII~$\lambda$9531 & \SiVI~1.965$\mu$m & \SIII/ & \SiX~1.430$\mu$m & \SIII/ & Pa$\beta$ & \SIII/ \\
Date & (erg/s/cm$^2$) & (erg/s/cm$^2$) & \SiVI & (erg/s/cm$^2$) & \SiX & (erg/s/cm$^2$) & Pa$\beta$ \\
\hline
2002 Apr 23 & (7.61$\pm$0.09)e$-$13 & (6.75$\pm$0.23)e$-$14 & 11.3$\pm$0.4 & (4.75$\pm$0.22)e$-$14 & 16.0$\pm$0.8 & (9.22$\pm$0.09)e$-$13 & 0.825$\pm$0.013 \\
2004 May 23 & (1.06$\pm$0.02)e$-$12 & (9.91$\pm$0.51)e$-$14 & 10.7$\pm$0.6 & (6.23$\pm$0.58)e$-$14 & 17.0$\pm$1.6 & (7.59$\pm$0.24)e$-$13 & 1.397$\pm$0.053 \\
2006 Jan  8 & (1.17$\pm$0.01)e$-$12 & (8.56$\pm$0.33)e$-$14 & 13.7$\pm$0.5 & (4.51$\pm$0.19)e$-$14 & 25.9$\pm$1.1 & (1.11$\pm$0.01)e$-$12 & 1.054$\pm$0.014 \\
2006 Jun 12 & (9.97$\pm$0.10)e$-$13 & (7.59$\pm$0.20)e$-$14 & 13.1$\pm$0.4 & (6.26$\pm$0.50)e$-$14 & 15.9$\pm$1.3 & (9.19$\pm$0.13)e$-$13 & 1.085$\pm$0.018 \\
2007 Jan 24 & (8.17$\pm$0.14)e$-$13 & (6.58$\pm$0.18)e$-$14 & 12.4$\pm$0.4 & (4.18$\pm$0.24)e$-$14 & 19.6$\pm$1.2 & (8.04$\pm$0.26)e$-$13 & 1.016$\pm$0.037 \\
2010 Feb 27 & (8.89$\pm$0.15)e$-$13 & (7.72$\pm$0.68)e$-$14 & 11.5$\pm$1.0 & (5.79$\pm$0.77)e$-$14 & 15.4$\pm$2.1 & (1.64$\pm$0.01)e$-$12 & 0.542$\pm$0.010 \\
\hline
\end{tabular}
\end{table*}

\begin{table*}
\caption{\label{optlines} 
Optical emission line fluxes and ratios}
\begin{tabular}{lccccccccc}
\hline
Observation & \OIII~$\lambda$5007 & \OIII~$\lambda$4959 & \OIII/ & \FeVII~$\lambda$3759 & \OIII/ & \FeVII~$\lambda$5159 & \OIII/ \\
Date & (erg/s/cm$^2$) & (erg/s/cm$^2$) & \OIII & (erg/s/cm$^2$) & \FeVII & (erg/s/cm$^2$) & \FeVII \\
\hline
2002 Apr 11 & (8.52$\pm$0.03)e$-$12 & (2.96$\pm$0.03)e$-$12 & 2.88$\pm$0.03 & (7.87$\pm$1.38)e$-$14 & 108.3$\pm$19.0 & (4.23$\pm$0.57)e$-$14 & 201.4$\pm$27.0 \\
2004 May 28 & (1.18$\pm$0.01)e$-$11 & (4.60$\pm$0.04)e$-$12 & 2.57$\pm$0.02 & (1.85$\pm$0.25)e$-$13 &  63.8$\pm$8.7  & (6.60$\pm$0.73)e$-$14 & 178.8$\pm$19.7 \\
2006 Jan  5 & (1.17$\pm$0.01)e$-$11 & (3.92$\pm$0.02)e$-$12 & 2.99$\pm$0.02 & (9.66$\pm$0.58)e$-$14 & 121.1$\pm$7.3  & (5.36$\pm$0.21)e$-$14 & 218.3$\pm$8.7 \\
2006 May 29 & (9.80$\pm$0.04)e$-$12 & (3.86$\pm$0.03)e$-$12 & 2.54$\pm$0.02 & (9.25$\pm$1.36)e$-$14 & 106.0$\pm$15.6 & (5.34$\pm$0.60)e$-$14 & 183.5$\pm$20.7 \\
2007 Feb  9 & (1.46$\pm$0.01)e$-$11 & (5.09$\pm$0.04)e$-$12 & 2.87$\pm$0.03 & (7.45$\pm$1.20)e$-$14 & 196.0$\pm$31.6 & (5.32$\pm$0.40)e$-$14 & 274.4$\pm$20.9 \\
2010 Feb 18 & (9.79$\pm$0.03)e$-$12 & (3.14$\pm$0.02)e$-$12 & 3.12$\pm$0.03 & (6.82$\pm$1.15)e$-$14 & 143.6$\pm$24.3 & (3.46$\pm$0.29)e$-$14 & 283.0$\pm$24.1 \\
\hline
Observation & \OIII~$\lambda$5007 &  \FeVII~$\lambda$5721 & \OIII/ & \FeVII~$\lambda$6087 & \OIII/ & H$\alpha$ & \OIII/ \\
Date & (erg/s/cm$^2$) & (erg/s/cm$^2$) & \FeVII & (erg/s/cm$^2$) & \FeVII & (erg/s/cm$^2$) & H$\alpha$ \\
\hline
2002 Apr 11 & (8.52$\pm$0.03)e$-$12 &  (8.67$\pm$0.68)e$-$14 &  98.3$\pm$7.7  & (1.19$\pm$0.10)e$-$13 & 71.6$\pm$5.7 & (1.43$\pm$0.01)e$-$11 & 0.596$\pm$0.005 \\
2004 May 28 & (1.18$\pm$0.01)e$-$11 &  (9.75$\pm$0.81)e$-$14 & 121.0$\pm$10.1 & (1.91$\pm$0.11)e$-$13 & 61.8$\pm$3.6 & (1.47$\pm$0.01)e$-$11 & 0.803$\pm$0.005 \\
2006 Jan  5 & (1.17$\pm$0.01)e$-$11 &  (8.73$\pm$0.39)e$-$14 & 134.0$\pm$6.0  & (1.64$\pm$0.07)e$-$13 & 71.3$\pm$3.2 & (1.64$\pm$0.01)e$-$11 & 0.713$\pm$0.003 \\
2006 May 29 & (9.80$\pm$0.04)e$-$12 &  (8.32$\pm$0.68)e$-$14 & 117.8$\pm$9.7  & (1.26$\pm$0.11)e$-$13 & 77.8$\pm$6.8 & (1.46$\pm$0.01)e$-$11 & 0.671$\pm$0.005 \\
2007 Feb  9 & (1.46$\pm$0.01)e$-$11 &  (9.93$\pm$0.62)e$-$14 & 147.0$\pm$9.1  & (1.73$\pm$0.08)e$-$13 & 84.4$\pm$4.1 & (1.41$\pm$0.01)e$-$11 & 1.035$\pm$0.007 \\
2010 Feb 18 & (9.79$\pm$0.03)e$-$12 &  (7.25$\pm$0.56)e$-$14 & 135.0$\pm$10.4 & (1.39$\pm$0.09)e$-$13 & 70.4$\pm$4.4 & (1.89$\pm$0.02)e$-$11 & 0.518$\pm$0.005 \\
\hline
\end{tabular}
\end{table*}

\begin{table*}
\caption{\label{ircont} 
Near-IR continuum fluxes and optical coronal line ratios}
\begin{tabular}{lccccccc}
\hline
Observation & $\lambda f_{1\mu{\rm m}}$ & $\lambda f_{1\mu{\rm m}}$/ & $\lambda f_{2.1\mu{\rm m}}$ & $\lambda f_{2.1\mu{\rm m}}$/ & $T_{\rm dust}$ & \FeVII~$\lambda$6087/ & \FeVII~$\lambda$5159/ \\
Date & (erg/s/cm$^2$) & \SIII & (erg/s/cm$^2$) & \SIII & (K) & \FeVII~$\lambda$3759 & \FeVII~$\lambda$6087 \\
\hline
2002 Apr 23 & (8.57$\pm$0.03)e$-$11 & 112.6$\pm$1.5 & (1.22$\pm$0.01)e$-$10 & 160.3$\pm$1.9 & 1316 & 1.512$\pm$0.290 & 0.355$\pm$0.055 \\
2004 May 23 & (7.79$\pm$0.25)e$-$11 &  73.5$\pm$2.9 & (1.18$\pm$0.01)e$-$10 & 111.3$\pm$2.7 & 1281 & 1.032$\pm$0.153 & 0.346$\pm$0.043 \\
2006 Jan  8 & (9.34$\pm$0.07)e$-$11 &  79.8$\pm$0.7 & (1.62$\pm$0.01)e$-$10 & 138.5$\pm$1.0 & 1328 & 1.698$\pm$0.127 & 0.327$\pm$0.020 \\
2006 Jun 12 & (6.67$\pm$0.07)e$-$11 &  66.9$\pm$1.0 & (1.50$\pm$0.01)e$-$10 & 150.5$\pm$1.7 & 1278 & 1.362$\pm$0.233 & 0.424$\pm$0.061 \\
2007 Jan 24 & (5.12$\pm$0.28)e$-$11 &  62.7$\pm$3.6 & (1.04$\pm$0.01)e$-$10 & 127.3$\pm$2.4 & 1231 & 2.322$\pm$0.390 & 0.308$\pm$0.028 \\
2010 Feb 27 & (2.18$\pm$0.01)e$-$10 & 245.2$\pm$4.4 & (3.47$\pm$0.01)e$-$10 & 390.3$\pm$6.6 & 1393 & 2.038$\pm$0.367 & 0.249$\pm$0.026 \\
\hline
\end{tabular}
\end{table*}

We have six epochs of quasi-simultaneous (within 3-14~days) near-IR
and optical spectroscopy for NGC~4151 (see Tables \ref{irlines} and
\ref{optlines}). The near-IR spectroscopy was obtained with the SpeX
spectrograph \citep{Ray03} at the NASA Infrared Telescope Facility
(IRTF), a 3 m telescope on Mauna Kea, Hawai'i, in the short
cross-dispersed mode (SXD, $0.8-2.4$ $\mu$m). All data except those
from 2010 were obtained through a slit of $0.8\times15''$ giving an
average spectral resolution of full width at half maximum (FWHM) $\sim
400$~km~s$^{-1}$. A narrower slit of $0.3\times15''$ was used for the
2010 epoch. The four epochs spanning the years $2004 - 2007$ are our
own data and were presented in \citet{L08a} and \citet{L11a}. The
near-IR spectra from 2002 and 2010 were discussed by \citet{Rif06} and
\citet{Schnuelle13}, respectively. The optical spectra were obtained
with the FAST spectrograph \citep{Fast98} at the Tillinghast \mbox{1.5
  m} telescope on Mt. Hopkins, Arizona, using the 300~l/mm grating and
a $3''$ long-slit. This set-up resulted in a wavelength coverage of
$\sim 3720-7515$~\AA~and an average spectral resolution of FWHM $\sim
330$~km~s$^{-1}$. The slit was rotated to the parallactic angle only
for the January 2006 epoch, however, except for the 2004 data, all
spectra were observed at a very low airmass ($\sec~z \sim 1.05$). The
May 2004 spectrum was observed at an airmass of $\sec~z \sim 1.3$ and
so the flux loss due to atmospheric differential refraction is
expected to be $\sim 20\%$ at the observed wavelength of
\FeVII~$\lambda$3759 relative to that at wavelengths
$\ge5000$~\AA~\citep{Fil82}. The January 2006 data were discussed in
\citet{L08a}, whereas all other optical spectra were retrieved from
the FAST archive.

We have measured the fluxes of the strongest near-IR and optical
coronal lines by integrating the observed profiles over the local
continuum, i.e. we have not assumed a specific line shape. In the
near-IR, we have measured two sulfur lines and two silicon lines,
namely, \SVIII~$\lambda$9911, \SIX~1.252$\mu$m, \SiVI~1.965$\mu$m and
\SiX~1.430$\mu$m (see Table \ref{irlines}). Two of these emission
lines are blended, namely, \SIX~with \FeII~1.257$\mu$m, and \SiVI~with
H$_2$~1.957$\mu$m. In these cases, we have assumed a Gaussian profile
in the deblending procedure. In the optical, we have measured four
iron emission lines, namely, \FeVII~$\lambda$3759,
\FeVII~$\lambda$5159, \FeVII~$\lambda$5721 and \FeVII~$\lambda$6087
(see Table \ref{optlines}). We have considered also
\FeX~$\lambda$6375, but this line is extremely weak in NGC~4151 and
makes up only $\sim 15\%$ of the total blend with \OI~$\lambda$6364
\citep{Pelat87}. We do not observe significant variability in the flux
of the entire blend, which constrains the \FeX~variability to below a
factor of $\sim 2$.

Since both the near-IR and optical spectra were obtained in
non-photometric sky conditions, we study in the following the temporal
changes of the coronal lines in relative rather than absolute flux. In
particular, we scale the coronal line emission to that of a strong,
forbidden {\it low-ionisation} emission line that is unblended and
observed in the same spectrum. The emission region that produces the
low-ionisation narrow lines is believed to be located at large enough
distances from the central ionising source for its flux to remain
constant on timescales of decades. We have scaled the near-IR and
optical coronal lines to the \SIII~$\lambda$9531 and
\OIII~$\lambda$5007 emission lines, respectively.

The error estimates on the line fluxes are crucial for assessing the
significance of the coronal line variability. The data are of
relatively high signal-to-noise ratio (continuum $S/N \ga 50-100$)
and, therefore, the main sources of measurement errors are of a
subjective nature, namely, the placement of the local continuum and
related to this the setting of the extension of the emission
line. These are problematic in most cases since the lines in question
sit on top of broad emission lines, namely, \SIII~$\lambda$9531 in the
blue wing of Pa$\epsilon$, \SVIII~in the blue wing of Pa$\delta$,
\SIX~in the blue wing of Pa$\beta$, \SiVI~in the red wing of
Br$\delta$, \OIII~$\lambda$5007 in the red wing of H$\beta$
and~\FeX~in the blue wing of H$\alpha$. Therefore, in order to
estimate meaningful uncertainties for the measured line fluxes we have
considered in addition to the statistical errors due to the data
quality also profile comparisons. These were done in velocity space
between the scaling lines \SIII~$\lambda$9531 and \OIII~$\lambda$5007
and the coronal lines and helped us judge the influence of the
continuum placement on the recovery of the true line profile. The
total estimated $1\sigma$ uncertainties for all measured line fluxes
are listed in Tables \ref{irlines} and \ref{optlines}. They are $\sim
1-2\%$ for the strongest lines and $\sim 3-15\%$ for the weakest ones.

In order to further constrain the significance of the observed coronal
line variability, we have considered the two extreme cases of where no
variability and the highest variability are expected. For both scaling
lines we observe also the other emission line that is emitted from the
same upper level, namely, \SIII~$\lambda 9069$ and \OIII~$\lambda
4959$.  Their observed ratios, which should be close to the
theoretical values of \SIII~$\lambda 9531$/\SIII~$\lambda 9069$=2.58
and \OIII~$\lambda 5007$/\OIII~$\lambda 4959$=2.92 \citep{NIST}, are
not expected to vary and so their observed variability sets a lower
threshold for the significance of the coronal line variability. Then,
we have measured the fluxes of the two prominent broad emission lines
Pa$\beta$ (in the near-IR) and H$\alpha$ (in the optical). Since the
BELR is expected to be the most variable of any AGN emission line
region, the observed variability of these broad lines gives an
estimate of the maximum value that can be reached within the current
data set. These results are also listed in Tables \ref{irlines} and
\ref{optlines}.

In addition to the emission lines, we have measured in the near-IR
spectra the continuum fluxes at the rest-frame wavelengths of $\sim
1~\mu$m and $\sim 2.1~\mu$m (see Table \ref{ircont}). As we have shown
in \citet{L11a}, the former is dominated by the accretion disc flux,
which is believed to be the main source of ionising radiation in AGN
and so the driver of the observed variability, whereas the latter is
emitted from the hot dust component of the obscuring torus, which, if
its location indeed coincides with that of the coronal line region,
should have a variability response similar to it. Furthermore, we have
derived the hot dust temperature from blackbody fits to the near-IR
spectral continuum as described in \citet{L11a} and list it also in
Table \ref{ircont}. In Section \ref{location}, we will compare these
values to the optical coronal line ratios
\FeVII~$\lambda$6087/\FeVII~$\lambda$3759 and
\FeVII~$\lambda$5159/\FeVII~$\lambda$6087, which are suitable
indicators of the gas temperature and density \citep{Nuss82, Keen87},
respectively (listed in Table \ref{ircont}).

\subsection{X-ray spectroscopy}

\begin{table}
\caption{\label{xraylog}
XMM-{\it Newton} archival observations}
\begin{tabular}{clrcl}
\hline
No.       & Observation          & Exp. & Obs ID      & Campaign \\
          & date and time        & (ks) &             & \\
\hline
 1        & 2000:12:21 16:45:30  &  33  & 0112310101  & 2000 Dec \\
 2$^\star$& 2000:12:21 14:21:13  &   8  & 0112310501  & \\
 3        & 2000:12:22 10:42:18  &  62  & 0112830201  & \\
 4        & 2000:12:22 02:53:42  &  23  & 0112830501  & \\
 5$^\star$& 2000:12:22 04:02:18  &   5  & 0112830601  & \\
 6        & 2003:05:25 08:37:56  &  19  & 0143500101  & 2003 May \\
 7        & 2003:05:26 20:35:02  &  19  & 0143500201  & \\
 8        & 2003:05:27 15:16:51  &  19  & 0143500301  & \\
 9        & 2006:05:16 06:21:42  &  40  & 0402660101  & 2006 May \\
10        & 2006:11:29 17:20:13  &  53  & 0402660201  & 2006 Nov \\
11        & 2011:05:11 15:20:09  &  16  & 0657840101  & 2011 May \\
12        & 2011:06:12 13:36:34  &  16  & 0657840201  & \\
\hline
\end{tabular}

\parbox[]{11cm}{$^\star$ no EPIC data}

\end{table}

Between December 2000 and June 2011, the time period that overlaps
with our optical and near-IR observations, there were five
observational campaigns with XMM-{\it Newton} on NGC~4151 (see Table
\ref{xraylog}). We have reduced these archival data sets using the
standard tasks in the XMM-{\it Newton} SAS software (version
13.5.). All the X-ray spectral analysis was done using the {\tt SPEX}
software\footnote{http://www.sron.nl/spex} \citep{Kaastra96}. We have
assumed solar abundances as given by \citet{Lod09} and corrected all
spectra for a Galactic column density of
2$\times$10$^{24}$~m$^2$. Three of the campaigns have multiple
observations, with two of them (2000 December and 2003 May) extending
over a period of 2-3 days and another one (2011 May) spanning a period
of a month. Since the X-ray emission lines are not expected to vary on
days to weeks timescales \citep{Det08, Det09} and in order to improve
the $S/N$ ratio, we have stacked the individual RGS spectra into one
spectrum per campaign. We have treated the May 16 and November 29,
2006, observations separately, since the time span between them is
about half a year. The longest and shortest observations are for the
December 2000 and May 2011 campaigns with total exposure times of
131~ks and 32~ks, respectively. 

In the X-rays, NGC~4151 is classified as an obscured Seyfert~1 galaxy,
i.e. the soft X-ray emission is heavily obscured by an
absorber. However, this absorption is not as deep as that observed for
Seyfert~2 galaxies and a very weak continuum is present. This absorber
is different from the warm absorber detected in the ultraviolet (UV)
and X-ray spectra of $\sim$50\% of Seyfert~1 galaxies
\citep{Cren03}. It has a much larger column density, outflow velocity
and velocity broadening, is variable on much shorter timescales and
only partially covers the X-ray continuum source. In order to
differentiate this absorber from the warm absorber, we follow
\citet{Kaastra14} and will call it 'the obscurer'. Since the soft and
hard X-ray continuua as well as the properties of the obscurer, such
as its column density, covering factor and potentially also the
ionisation, are known to be variable on day timescales or less, we
have not combined the EPIC pn spectra, but fitted them separately.

\begin{table*}
\caption{\label{xraylines} 
X-ray emission line and continuum fluxes}
\begin{tabular}{lccccc}
\hline
Observation & \OVII~f~0.561~keV & \OVII~i~0.569~keV & \OVII~r~0.574~keV & \OVIII~$\alpha$~0.654~keV & \NeIX~f~0.905~keV \\
Epoch & (ph/s/cm$^2$) & (ph/s/cm$^2$) & (ph/s/cm$^2$) & (ph/s/cm$^2$) & (ph/s/cm$^2$) \\
\hline
2000 Dec & (5.05$\pm$0.17)e$-$04 & (0.70$\pm$0.09)e$-$04 & (1.51$\pm$0.10)e$-$04 & (1.47$\pm$0.06)e$-$04 & (0.59$\pm$0.04)e$-$04 \\
2003 May & (5.10$\pm$0.21)e$-$04 & (0.43$\pm$0.09)e$-$04 & (1.36$\pm$0.13)e$-$04 & (2.02$\pm$0.06)e$-$04 & (0.56$\pm$0.04)e$-$04 \\
2006 May & (4.70$\pm$0.21)e$-$04 & (0.78$\pm$0.11)e$-$04 & (1.55$\pm$0.13)e$-$04 & (1.59$\pm$0.08)e$-$04 & (0.45$\pm$0.08)e$-$04 \\
2006 Nov & (3.89$\pm$0.24)e$-$04 & (0.55$\pm$0.31)e$-$04 & (1.49$\pm$0.34)e$-$04 & (1.60$\pm$0.08)e$-$04 & (0.49$\pm$0.07)e$-$04 \\
2011 May & (5.63$\pm$0.95)e$-$04 & (0.98$\pm$0.58)e$-$04 & (1.80$\pm$0.62)e$-$04 & (1.87$\pm$0.28)e$-$04 & (0.87$\pm$0.16)e$-$04 \\
\hline
Observation & \NVI~f~0.420~keV & \NVII~$\alpha$~0.500~keV & \CVI~$\alpha$~0.368~keV & $f_{\rm 2-10keV}$ \\
Epoch & (ph/s/cm$^2$) & (ph/s/cm$^2$) & (ph/s/cm$^2$) & (erg/s/cm$^2$) \\
\hline
2000 Dec & (1.51$\pm$0.09)e$-$04 & (0.87$\pm$0.05)e$-$04 & (2.66$\pm$0.13)e$-$04 & (6.37$\pm$0.06)e$-$11 \\
2003 May & (1.59$\pm$0.10)e$-$04 & (0.97$\pm$0.06)e$-$04 & (2.91$\pm$0.15)e$-$04 & (3.05$\pm$0.24)e$-$10 \\
2006 May & (1.27$\pm$0.13)e$-$04 & (0.94$\pm$0.08)e$-$04 & (3.10$\pm$0.22)e$-$04 & (7.91$\pm$0.08)e$-$11 \\
2006 Nov & (1.23$\pm$0.14)e$-$04 & (0.73$\pm$0.09)e$-$04 & (2.43$\pm$0.22)e$-$04 & (1.37$\pm$0.01)e$-$10 \\
2011 May & (1.05$\pm$0.48)e$-$04 & (0.84$\pm$0.28)e$-$04 & (2.63$\pm$0.74)e$-$04 & (1.55$\pm$0.05)e$-$10 \\
\hline
\end{tabular}
\end{table*}

The presence of the obscurer in NGC~4151 allows us to study in detail
the X-ray emission lines, which would otherwise be swamped by the high
continuum flux. However, the obscurer also complicates the analysis,
in particular the continuum fitting, since the absorption is so deep
that the individual absorption lines blend and create a
pseudo-continuum. Decomposing the total absorption into obscurer and
warm absorber and characterising each individually is not possible
without prior knowledge of the ionisation structure of the warm
absorber. This, however, is not known for NGC~4151 since it has never
been observed in an unobscured state. However, for the Seyfert~1
galaxy NGC~5548 we have high-quality X-ray spectra in both the
unobscured \citep{KS03, KS05, Det09} and obscured state
\citep{Kaastra14}. Thus, we have assumed in our X-ray analysis that
the obscurer in NGC~4151 has the same structure as that in NGC~5548,
albeit being deeper. This is supported by the fact that the
\CIV~absorption lines observed in the {\it Hubble Space Telescope}
(HST) UV spectra of NGC~4151 \citep{Kraemer06} and NGC~5548
\citep{Kaastra14} have a similar width and blueshift.

Following the approach chosen by \cite{Kaastra14} for NGC~5548, we
first fitted the May 2003 spectrum, since it is the least absorbed
state and allows for a better determination of the warm absorber and
obscurer components. Each warm absorber or obscurer component was
fitted with an {\tt xabs} model. Then, using their results for
NGC~5548, we assumed that the obscurer has two components, one close
to or neutral with a lower covering factor and another medium-ionised
with a much larger covering factor. From the UV spectrum of NGC~4151
we know that there are at least six warm absorber components with
different velocities \citep[see Fig. 2 in][]{Kraemer06}, but their
ionisation structure is undetermined by these data. We have added a
further four warm absorber components (each an {\tt xabs} model) with
different ionisation parameters. Although the ionisation parameter of
the warm absorber and obscurer can overlap, we only allowed higher
ionization parameters for the warm absorber. The free parameters in
the fit were the hydrogen column density and ionisation parameter, and
in the case of the obscurer also the covering factor. The outflow
velocity and velocity broadening were left at the standard values
since they cannot be constrained. Finally, we assumed that the narrow
emission lines are unaffected by the obscurer and warm absorber.

We modelled the continuum as observed with EPIC-pn with a reflection
component and a comptoionisation model, which dominate the hard
X-rays, and added a modified blackbody in the soft X-ray regime. The
fit also includes, but the parameters are frozen, the best fit model
for the emission lines, both narrow and broad, and the radiative
recombination continua as determined from the RGS spectra. The
Fe~K$\alpha$ and K$\beta$ lines, which can only be studied with the
EPIC instruments, were fitted as Gaussians. Both lines are
unresolved. We fit the spectra in the $0.3-10$~keV energy range and
used the option of optimal binning available in {\tt SPEX}. This gives
an acceptable fit to the EPIC-pn spectra. However, due to the heavy
absorption, there is a certain degeneracy between the normalisation
and temperature of the modified blackbody, the hydrogen column
density, ionisation parameter and for the obscurer also the covering
factor, of the absorption components and the normalisation of the
emission lines and radiative recombination continua (frozen in the
EPIC pn fit). In Table \ref{xraylines}, we list the fluxes and
$1\sigma$~errors for the eight strongest X-ray coronal lines, mainly
from oxygen but also from nitrogen, carbon and neon, and the
unabsorbed $2-10$~keV continuum fluxes. The errors on the line fluxes
are $\sim 3-10\%$ for the strongest lines and $\sim 10-20\%$ for the
weakest lines. The errors for the May 2011 observing campaign are much
larger on all measurements ($\sim 20-60\%$).

\section{The variability behaviour}

\subsection{The near-IR and optical coronal lines} \label{iroptvar}

\begin{figure}
\centerline{
\includegraphics[clip=true, bb=20 380 590 715, scale=0.45]{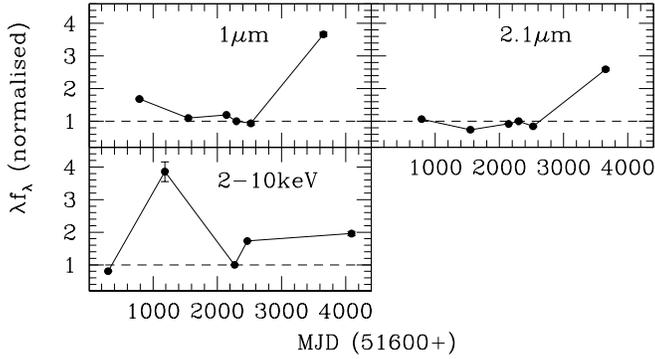}
}
\caption{\label{contvar} The variability of the continuum fluxes at
  rest-frame wavelengths of $\sim 1~\mu$m (sampling the accretion
  disc), $\sim 2.1~\mu$m (sampling the hot dust emission) and in the
  energy range $2-10$~keV (related to the accretion disc). The near-IR
  continuum fluxes were divided by the \SIII~$\lambda 9531$ line flux
  and all continuum fluxes were normalised to the value of the
  May/June 2006 epoch. We plot $1\sigma$ error bars.}
\end{figure}

\begin{figure}
\centerline{
\includegraphics[clip=true, bb=20 260 590 715, scale=0.45]{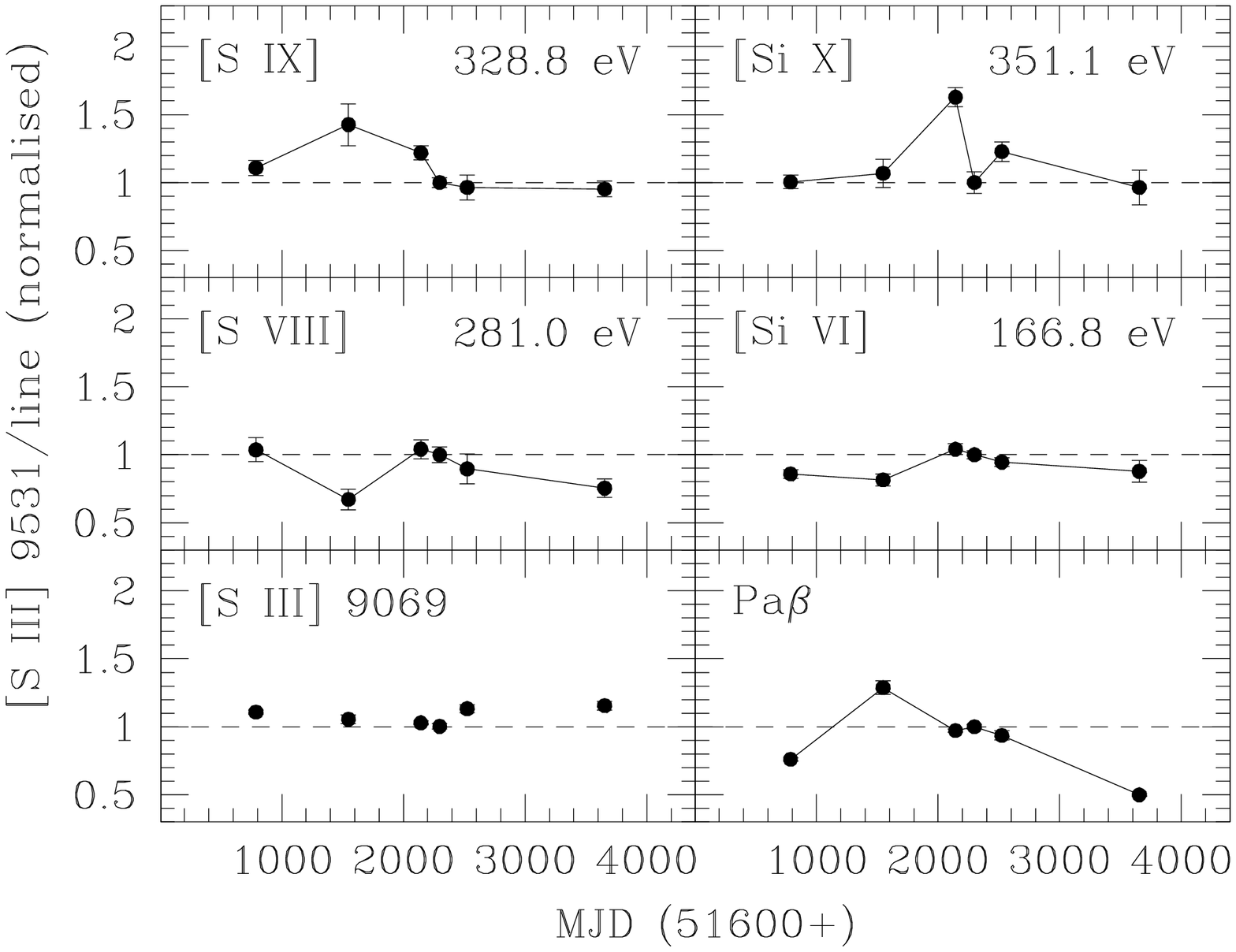}
}
\caption{\label{irvar} The variability of the near-IR coronal lines in
  the period April 2002 - February 2010. For comparison, we show also
  the \SIII~$\lambda 9069$ narrow line and Pa$\beta$ broad line, which
  are expected not to vary and to vary maximally, respectively. The
  \SIII~$\lambda 9531$/\SIII~$\lambda 9069$ flux ratios were
  normalised to the theoretical value, whereas those between
  \SIII~$\lambda 9531$ and the other lines were normalised to the
  value of the June 2006 epoch. We plot $1\sigma$ error bars. The
  ionisation potentials of the coronal lines are given at the top
  right.}
\end{figure}

\begin{figure}
\centerline{ 
\includegraphics[clip=true, bb=20 260 590 715, scale=0.45]{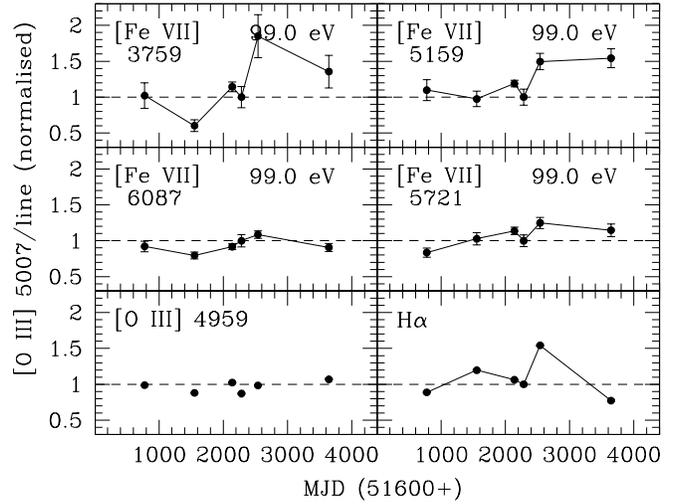}
}
\caption{\label{optvar} Same as Fig. \ref{irvar} for the optical
  coronal lines. For comparison, we show also the \OIII~$\lambda 4959$
  narrow line and H$\alpha$ broad line, which are expected not to vary
  and to vary maximally, respectively.}
\end{figure}

\begin{figure}
\centerline{ 
\includegraphics[scale=0.45]{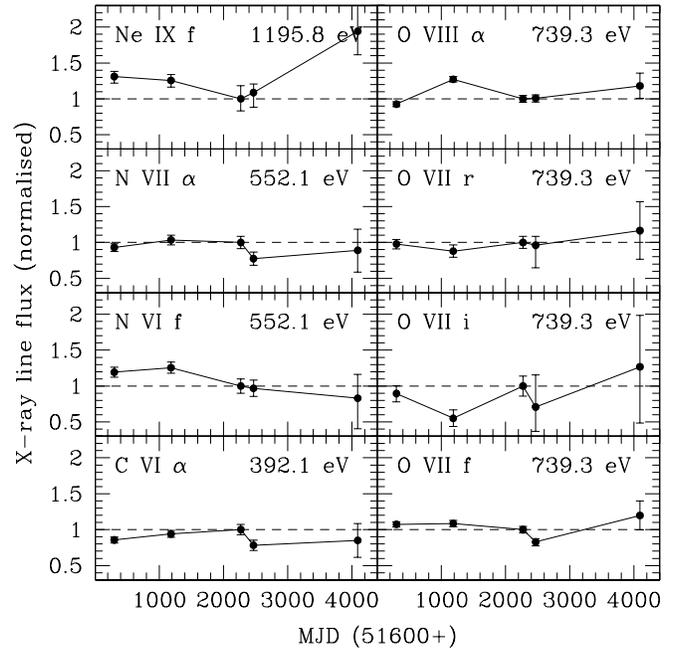}
}
\caption{\label{xrayvar} The variability of the X-ray coronal lines in
  the period December 2000 - May 2011. The flux values were normalised
  to the value of the May 2006 epoch. We plot $1\sigma$ error
  bars. The ionisation potentials of the coronal lines are given at
  the top right.}
\end{figure}

In the $\sim 8$~year period sampled by the near-IR and optical
spectroscopy the ionising flux of the accretion disc, which is assumed
to be the main driver for the variability of the broad line, coronal
line and hot dust emission regions, decreased by a factor of $\sim 2$
in the first five years and increased by a factor of $\sim 4$ in the
following three years (see Fig. \ref{contvar}). We note that the flux
increase of $\sim 9\%$ observed between the May 2004 and January 2006
epochs is consistent with zero once the deviations from the
theoretical value of the \SIII~$\lambda 9531$/\SIII~$\lambda 9069$
ratios measured in these spectra are taken into account ($\sim 5.5\%$
and $\sim 2.9\%$, respectively). The Pa$\beta$ broad line region and
hot dust emission responded to the accretion disc variability in a
similar way (see Figs. \ref{irvar} and \ref{contvar}, respectively);
their flux decreased in the first two years by $\sim 70\%$ and $\sim
50\%$, respectively, and then continuously increased for the next six
years by a factor of $\sim 3$ and $\sim 4$, respectively. We note that
the flux decrease of $\sim 20\%$ observed for the hot dust emission
between the June 2006 and January 2007 epochs is similar to the
deviation from the theoretical value measured for the \SIII~$\lambda
9531$/\SIII~$\lambda 9069$ ratio in the latter spectrum ($\sim
13.3\%$). The flux of the H$\alpha$ broad line region also decreased
in the first two years (by $\sim 35\%$) and increased in the next two
years (by $\sim 20\%$), but then it decreased for a short period
(between June 2006 and January 2007 by $\sim 54\%$) before it
increased strongly for the last three years (by a factor of $\sim 2$;
see Fig. \ref{optvar}). The intermittent flux decrease might be caused
by a reddening event, which we discuss further in Section
\ref{location}.

The variability response of the near-IR coronal lines in the four year
period between January 2006 and February 2010 is similar to that of
the Pa$\beta$ broad line region and hot dust emission, with a trend
that the higher the ionisation potential, the higher the flux
increase, namely, $\sim 69\%$ for the \SiX~line, $\sim 28\%$ for the
\SIX~line, $\sim 38\%$ for the \SVIII~line, and no significant
variability for the \SiVI~line (see Fig. \ref{irvar}). In the first
two years sampled by the data significant variability is observed only
for the \SVIII~line; a flux {\it increase} by $\sim 54\%$. Then, for
all but the \SIX~line, a flux decrease (instead of the increase seen
for the broad line region and hot dust emission) is observed between
May 2004 and January 2006, namely, $\sim 52\%$ for the \SiX~line,
$\sim 55\%$ for the \SVIII~line, and $\sim 28\%$ for the \SiVI~line.

Between April 2002 and January 2006 we detect significant variability
for only one optical coronal line, namely, \FeVII~$\lambda 3759$,
which behaves similar to the \SVIII~line (see Fig. \ref{optvar}); a
flux increase (by $\sim 70\%$) followed by a flux decrease (by $\sim
90\%$). The flux decrease observed for the H$\alpha$ broad line region
between June 2006 and January 2007 is clearly evident for the
\FeVII~$\lambda 3759$, \FeVII~$\lambda 5159$ and \FeVII~$\lambda
5721$~lines; their fluxes changed by $\sim 85\%$, $\sim 50\%$ and
$\sim 25\%$, respectively. If we consider the period of January 2006
to January 2007 instead, we observe a similar behaviour also for the
\FeVII~$\lambda 6087$~line; its flux decreased by $\sim 18\%$. In the
following three years the \FeVII~$\lambda 6087$~line flux increased by
$\sim 20\%$, as observed for the broad line region and hot dust
emission, whereas none of the other optical coronal lines varied
significantly.

\subsection{The X-ray coronal lines}

In the $\sim 10.5$~year period sampled by the X-ray spectroscopy the
unabsorbed $2-10$~keV continuum flux, which is assumed to be produced
by the central ionising source and so to be linked to the accretion
disc flux, increased by a factor of $\sim 5$ in the first two and a
half years, decreased by a factor of $\sim 4$ in the next three years,
and then increased again by a factor of $\sim 2$ in the next five
years.

The variability response of the three forbidden X-ray coronal lines in
the period between November 2006 and May 2011 is similar to that of
the near-IR coronal lines and that of the Pa$\beta$ broad line region
and hot dust emission, and again with a trend that the higher the
ionisation potential, the higher the flux increase, namely, a factor
of $\sim 2$ for the \NeIX~f~line, $\sim 45\%$ for the \OVII~f~line and
no significant variability for the \NVI~f line (see
Fig. \ref{xrayvar}). Then, whereas the flux stayed roughly constant in
the first two and half years sampled by the data, similar to the
near-IR coronal lines, a flux decrease (instead of the increase seen
for the broad line region and hot dust emission) is observed in the
period between May 2003 and May 2006 for the \OVII~f~line (by $\sim
31\%$) and the \NVI~f~line (by $\sim 30\%$).

In the entire time period sampled by the data no significant
variability is observed for the recombination line \OVII~r, whereas
the intercombination line \OVII~i showed strong variability between
December 2000 and May 2006, with a flux decrease by $\sim 62\%$
followed by a flux increase by $\sim 82\%$. No significant variability
is observed for the \OVII~i~line after May 2006, which could be due to
the relatively large errors on the line flux for the following two
observing epochs. The \OVIII~$\alpha$~line showed also significant
variability between December 2000 and May 2006, but contrary to the
\OVII~i~line, its flux increased first (by $\sim 37\%$) and then
decreased (by $\sim 27\%$). Also for this line there is no significant
variability after May 2006. The variability response of the two
permitted X-ray coronal lines with the highest ionisation potentials
is similar to that of the \OVII~f~line; a significant flux decrease is
observed for the \NVII~$\alpha$~line only between May 2003 and
November 2006 (by $\sim 34\%$) and for the \CVI~$\alpha$ line only
between May 2006 and November 2006 (by $\sim 28\%$).

\section{The origin of the coronal line emission region} \label{location}

The current understanding is that the coronal line region in AGN is
dust-free, photoionised and located beyond the BELR at distances from
the central ionising source similar to those of the hot inner face of
the obscuring dusty torus. In this scenario, we expect to observe the
coronal line flux to vary similarly to that of the broad lines and hot
dust emission but on timescales longer and shorter than the former and
latter, respectively. With our data set we can test the first premise,
although its sparse time sampling does not allow us to measure any
variability timescales.

The variability behaviour observed for the broad lines and hot dust
emission in the period covered by our data is both similar and simple;
a flux decrease of $\sim 40-70\%$ in the first two years followed by a
strong flux increase by a factor of $\sim 2-4$ in the following six
years. {\it This variability behaviour is not observed for any of the
  coronal lines.} In general, the coronal lines varied weakly if at
all, with the largest flux change observed to be only $\sim 50-90\%$
in a period of about two years. Specifically, in the first four years
sampled by the data, the coronal lines either did not vary
significantly or showed the opposite behaviour to that of the broad
lines and hot dust emission. In the last four years sampled by the
data, only the coronal lines with the highest ionisation potentials
showed a variability behaviour similar to that of the broad lines and
hot dust emission, but with a much reduced amplitude (a lower flux
change per year by a factor of $\sim 2-4$), whereas the flux of the
coronal lines with relatively low ionisation potentials remained
unchanged.

The characteristic variability response time is a sum of the light
travel time, which depends mainly on the location of the emission
region, and the recombination time, which depends strongly on the gas
number density:

\begin{equation}
\tau_{\rm var} = \tau_{\rm lt} + \tau_{\rm rec}
\end{equation}

\noindent
Therefore, the low variability amplitude observed for the coronal line
gas in NGC~4151 indicates either that it is located well beyond the
broad line region and dusty torus or that it has a relatively low
density or both. In the following, we use published results from
high-spatial resolution imaging campaigns at near-IR, optical and
X-ray frequencies and apply plasma diagnostics to our own data in
order to constrain both the location and gas number density of the
coronal line emission region.

\subsection{Published high-spatial resolution observations} \label{imaging}

NGC~4151 has a well-measured hydrogen broad line region lag time of
about a week \citep{Zu11} and a hot dust lag time varying between
$\sim 30-70$~days \citep{Kosh14}. One possible explanation for the
fact that the coronal line region reacted much weaker to changes in
the ionising flux than both the broad line and hot dust emission
regions is that it is located much further out than them.

High-spatial resolution observations of NGC~4151 in the near-IR were
presented by several authors. \citet{Storchi09} used the Gemini
Near-IR Integral Field Spectrograph (NIFS) on December 2006 to image
the source in the $0.94-2.42$~$\mu$m wavelength range with a spatial
resolution of $0.12-0.16$~arcsec, which corresponds to $\approx
8-10$~pc at the source. Their contour maps of the \SVIII~and
\SIX~coronal lines show a compact, spatially unresolved
region. \citet{Mueller11} and \citet{Iser13} used the Keck OH
Suppressing InfraRed Imaging Spectrograph (OSIRIS) on March 2006 and
February/May 2005, respectively, to image the source in the
$1.96-2.38$~$\mu$m wavelength range with a spatial resolution of
$0.08-0.11$~arcsec, which corresponds to $\approx 5-7$~pc at the
source. Their contour maps of the \SiVI~coronal line show a bright
nucleus containing $\sim 70\%$ of the total emission, which is
surrounded by low-level extended emission up to $\approx 50-80$~pc.

The source NGC~4151 was observed with the Space Telescope Imaging
Spectrograph (STIS) on-board the {\sl Hubble Space Telescope (HST)} on
several occasions. These observations span a large optical wavelength
range and have a spatial resolution of $0.1$~arcsec, corresponding to
$\approx 7$~pc at the source. \citet{Nelson00} presented measurements
of the \FeVII~$\lambda 3759$, \FeVII~$\lambda 5721$~and
\FeVII~$\lambda 6087$~coronal line fluxes along the slit at two
different position angles for observations taken between January and
June 1998. Along one of the position angles (P.A. $70^\circ$), the
\FeVII~$\lambda 3759$ emission was spatially unresolved, but the
\FeVII~$\lambda 5721$~and \FeVII~$\lambda 6087$~emission lines showed
extents of up to \mbox{$\approx 70-100$~pc} from the nucleus. Along
the other position angle (P.A. $221^\circ$), all three \FeVII~emission
lines were spatially resolved, however, with much lesser extents of
$\approx 20-50$~pc from the nucleus. In all cases, the total
\FeVII~emission was dominated by the nuclear flux.

\citet{Wang11a, Wang11b} analysed the deep exposure of NGC~4151 taken
in March 2008 with the Advanced CCD Imaging Spectrometer (ACIS)
on-board the {\sl Chandra} X-ray observatory. By applying subpixel
event repositioning and binning techniques they were able to improve
the effective spatial resolution of the ACIS images to better than
0.4~arcsec, which corresponds to $\approx 25$~pc at the source. Their
emission line maps of \OVII~(f, i and r), \OVIII~$\alpha$ and
\NeIX~(f, i and r) show an extremely bright nucleus surrounded by
extended emission in the inner region of $\approx 130$~pc, and for the
\OVII~and \OVIII~emission lines also very low-level emission ($\sim
10\%$ of the total flux) up to $\approx 2$~kpc.

In summary, high-spatial resolution observations of NGC~4151 show that
whereas its coronal line emission region is extended in some chemical
species, the total flux that our variability study is most sensitive
to is dominated by the unresolved nuclear region. This region can
currently be constrained by direct imaging only to $\la 5$~pc (or $\la
15$~light years), which is much larger than the measured distance of
the hot dust emission of $\sim 2$ light months.

\subsection{Plasma diagnostics} \label{plasma}

\begin{figure}
\centerline{
\includegraphics[scale=0.43]{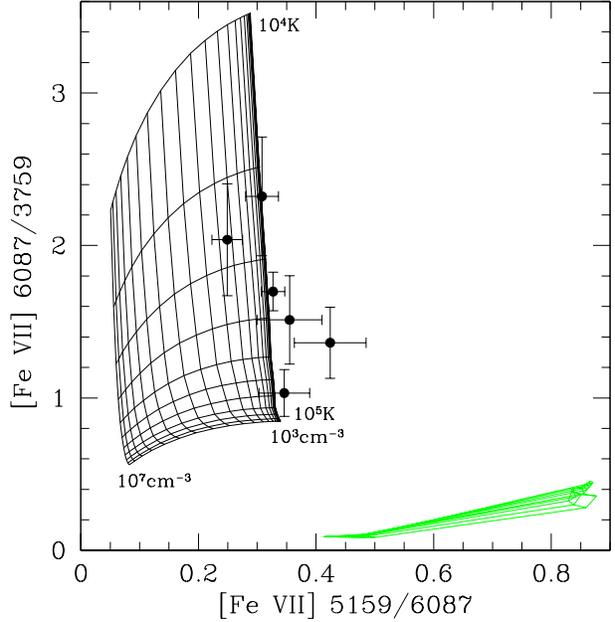}
}
\caption{\label{FeVIItemp} The observed optical coronal line ratios
  \FeVII~$\lambda$6087/$\lambda$3759 versus
  \FeVII~$\lambda$5159/$\lambda$6087, overlaid with curves of constant
  temperature ($\log T=4, 4.1, 4.2, ... 5$~K) and constant number
  density ($\log n_e=3, 3.2, 3.4, ... 7$~cm$^{-3}$) for the case of
  photoionisation equilibrium. The case of collisional ionisation
  equilibrium is shown in green. We plot $1\sigma$ error bars.}
\end{figure}

\begin{figure}
\centerline{
\includegraphics[scale=0.43]{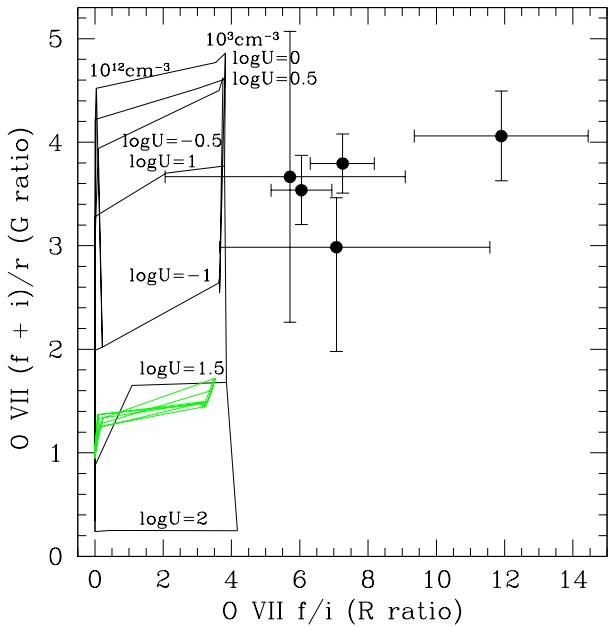}
}
\caption{\label{OVIItemp} The observed X-ray coronal line ratios
  \OVII~(f$+$i)/r (G ratio) versus \OVII~f/i (R ratio), overlaid with
  curves of constant ionisation parameter ($\log U=-1, -0.5, 0,
  ... 2$) for two constant number densities ($n_e=10^3$~cm$^{-3}$ and
  $n_e=10^{12}$~cm$^{-3}$) and a fixed column density of $N_{\rm
    H}=10^{21}$~cm$^{-2}$, assuming photoionisation equilibrium. The
  case of collisional ionisation equilibrium is shown in green. We
  plot $1\sigma$ error bars.}
\end{figure}

\begin{figure}
\centerline{
\includegraphics[clip=true, bb=12 426 425 840, scale=0.5]{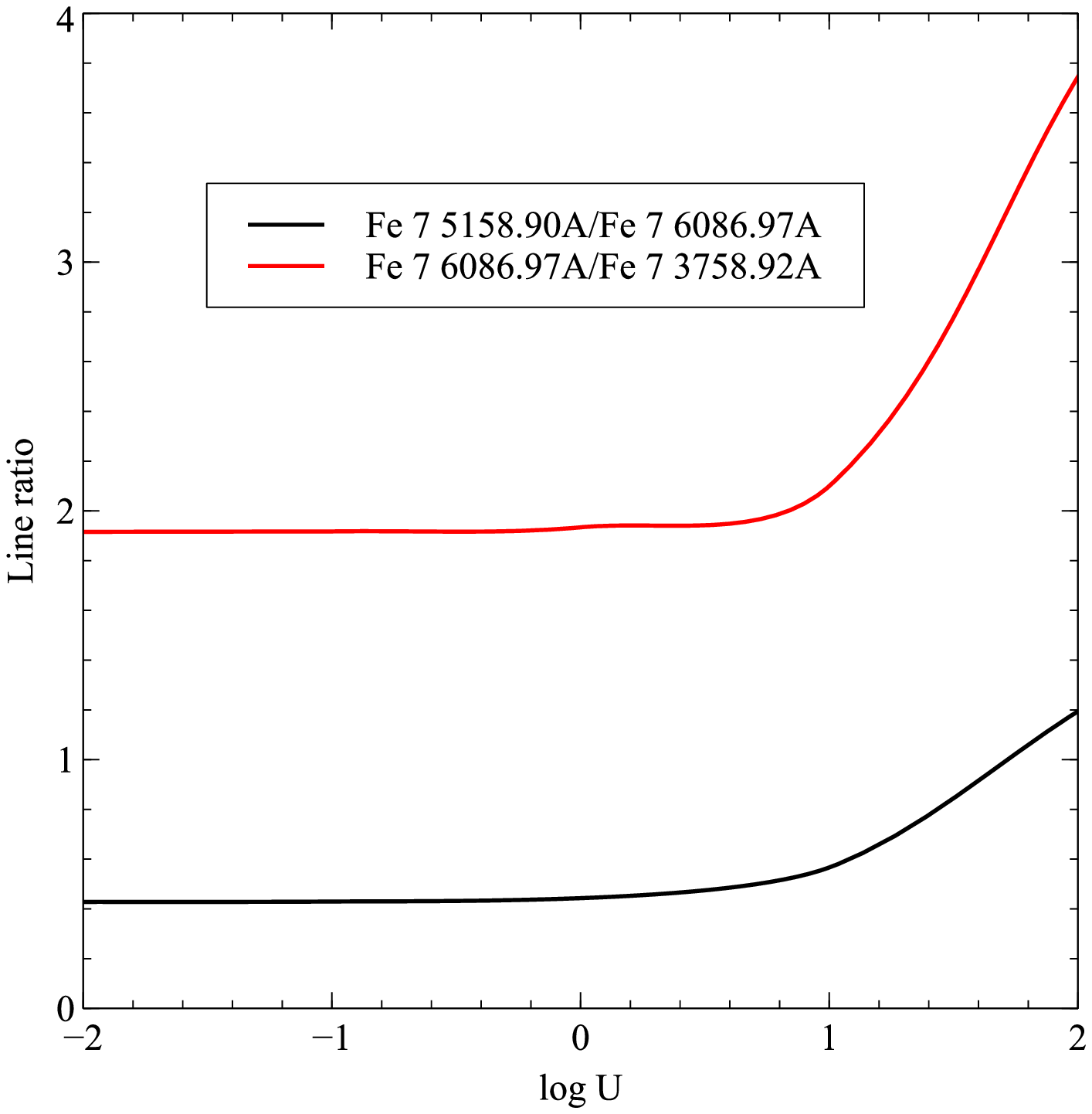}
}
\caption{\label{FeVIIpump} The effect of continuum pumping on the
  optical coronal line ratios \FeVII~$\lambda$6087/$\lambda$3759 (red
  solid line) and \FeVII~$\lambda$5159/$\lambda$6087 (black solid
  line) in dependence of the ionisation parameter. The gas is assumed
  to have a temperature of $T_e=10^{4.25}$~K and a number density of
  $n_e=10^3$~cm$^{-3}$.}
\end{figure}

\begin{figure}
\centerline{
\includegraphics[scale=0.4]{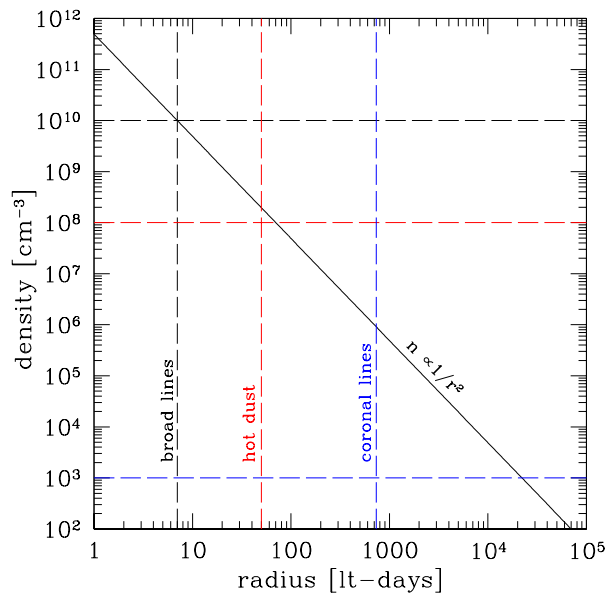}
}
\caption{\label{dens} The density run with radius calculated for the
  relationship $n \propto r^{-2}$ (black solid line) and scaled to the
  observed lag time and expected gas number density for the hydrogen
  broad line region in NGC 4151 (black dashed lines). The red dashed
  lines show the observed lag time and expected gas number density for
  the hot dust. The estimated density and distance of the coronal line
  region are indicated by the blue dashed lines.}
\end{figure}

\begin{figure}
\centerline{
\includegraphics[scale=0.42]{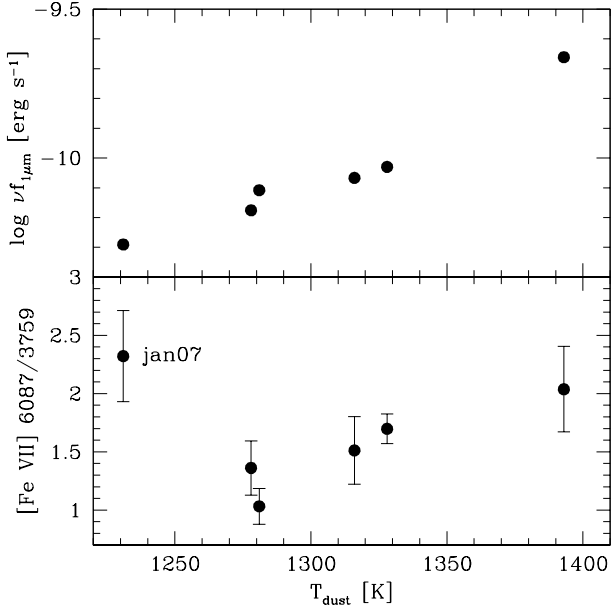}
}
\caption{\label{dusttemp} The top and bottom panels show the continuum
  flux at rest-frame wavelength of $\sim 1~\mu$m (sampling the
  accretion disc) and the optical coronal line ratio
  \FeVII~$\lambda$6087/$\lambda$3759, which is a suitable indicator of
  the gas temperature (the higher its value, the lower the
  temperature; see Fig. \ref{FeVIItemp}), respectively, versus the hot
  dust blackbody temperature.}
\end{figure}

For the optical and X-ray coronal line gas we can estimate the density
using the three iron lines \FeVII~$\lambda$3759, \FeVII~$\lambda$5159
and \FeVII~$\lambda$6087 and the three lines from the helium-like ion
of oxygen \OVII~f, \OVII~i and \OVII~r, respectively. Using the plasma
simulation code \cloudy~\citep[last described by][]{Cloudy}, we have
generated a temperature versus density grid for the collisionally
excited \FeVII~lines (see Fig. \ref{FeVIItemp}) and a ionisation
parameter versus density grid for the \OVII~lines (see
Fig. \ref{OVIItemp}). The iron collision strengths were taken from
\citet{Witt08}. For the photoionisation simulations we have
approximated the incident radiation field with the mean AGN spectral
energy distribution derived by \citet{Math87}. As discussed by
previous studies, the line ratios \FeVII~$\lambda$6087/$\lambda$3759
and \FeVII~$\lambda$5159/$\lambda$6087 are suitable indicators of
temperature and density, respectively \citep{Nuss82, Keen87}, whereas
the X-ray line ratios \OVII~(f$+$i)/r (the so-called G ratio) and
\OVII~f/i (the so-called R ratio) trace the ionisation parameter,
which is directly related to the kinetic gas temperature, and density,
respectively, for a given column density \citep{Por00, Porter07}. We
assumed the cases of either photoionisation or collisional ionisation
equilibrium.

The measurements for NGC~4151 in both Figs. \ref{FeVIItemp} and
\ref{OVIItemp} give two main consistent results; the coronal line gas
is photoionised rather than collisionally ionised and its density
appears to be relatively low. In the case of the optical coronal line
gas, if we assume that the gas density did not change with time, four
of the observing epochs constrain it to $n_e \sim 10^3$~cm$^{-3}$
within $\sim 1\sigma$, whereas the two observing epochs January and
June 2006 reach this value within $\sim 2\sigma$. If we assume the gas
density changed between observing epochs, at the $2 \sigma$~level it
varied between $n_e \sim 10^{5.2}$~cm$^{-3}$ and $n_e \la
10^3$~cm$^{-3}$, since at the low-density limit the grid contours fall
on top of each other. The highest density value is about two orders of
magnitude lower than the critical density of \FeVII~of $n_{\rm crit}
\sim 3 \times 10^7$~cm$^{-3}$. We get a similar result for the X-ray
coronal line gas; assuming the gas density did not change with time,
two observing epochs (November 2006 and May 2011) constrain it to $n_e
\sim 10^3$~cm$^{-3}$ within $\sim 1\sigma$, whereas three observing
epochs reach this value within $\sim 3\sigma$. If we assume the
density changed between observing epochs, at the $2 \sigma$~level it
varied between $n_e \sim 10^{12}$~cm$^{-3}$ (i.e. close to the
critical density) and $n_e \la 10^3$~cm$^{-3}$, since at the
low-density limit the grid contours fall on top of each other. 

It is worth noting that all measured \OVII~ratios and most of the
\FeVII~ratios are to the right of the theoretical grids. In the case
of \OVII, there are no known uncertainties in the atomic data and
adding a contribution from collisionally ionised plasma would not
increase the R ratios, which are observed to be much higher than
predicted by theory. Similarly high R ratios are found also in other
AGN, e.g., ${\rm R}=5.50$ for Mrk~3 \citep{Pounds05} and ${\rm
  R}=7.73$ for NGC~1068 \citep{Kraemer15}. However, in the case of
\FeVII, it is possible that some contribution from fluorescence is
present given the relatively low gas density in combination with the
large ionisation parameter we estimate below. The large number of
permitted \FeVIIp~lines in the far-UV can be pumped by the continuum,
which, when radiatively decaying, will add to the lower levels
responsible for the \FeVII~lines studied here. Fig. \ref{FeVIIpump}
shows the importance of this effect for a gas of number density
$n_e=10^3$~cm$^{-3}$ in dependence of the ionisation parameter. At a
ionisation parameter of $\log U \sim 1$ the expected increase in the
\FeVII~$\lambda$5159/$\lambda$6087 ratio is a factor of $\sim 1.5$,
whereas the increase in the \FeVII~$\lambda$6087/$\lambda$3759 ratio
is only $\sim 10\%$.

From Fig. \ref{FeVIItemp} we estimate a temperature of $T \sim
10^{4.25}$~K$\sim 18000$~K for the Fe$^{6+}$ gas. At this temperature,
the recombination time is:

\begin{equation}
\tau_{\rm rec} \approx \frac{1}{n_e \cdot 1.70 \times 10^{-10}}~~{\rm s},
\end{equation} 

\noindent
which for a gas number density of $n_e=10^3$~cm$^{-3}$ results in
$\tau_{\rm rec} \approx 5.9 \times 10^6$~s~$\approx 2.2$~months. We
note that we have used the total recombination coefficient, i.e. the
sum of the radiative and dielectronic recombination coefficients,
which was calculated based on the atomic data presented by
\citet{Gu06}. This value differs from that listed in
\citet{Osterbrock2} by about an order of magnitude, since the latter
represented only the radiative recombination coefficient. From
Fig. \ref{OVIItemp}, we constrain the ionisation parameter for all
X-ray observing epochs at the $1\sigma$ level to $\log U \sim
1$. Then, using the unabsorbed X-ray luminosity of $L_{\rm 2-10keV}
\sim 10^{43}$~erg~s$^{-1}$ resulting from our fits for the
highest-flux epoch as a proxy for the ionising luminosity producing
\OVII~and the best-fit X-ray spectral slope of this epoch of $\Gamma
\sim 1.43$ to estimate the mean ionising photon energy, the calculated
distance of the coronal line region from the central ionising source
for the above gas density is $\tau_{\rm lt} \sim 730$~light days~$\sim
2.0$~light years. This value, which is far below the resolution
provided by current direct imaging results (see Section
\ref{imaging}), puts this region well beyond the hot inner face of the
obscuring dusty torus. Together our estimates of $\tau_{\rm rec}$ and
$\tau_{\rm lt}$ give a characteristic variability response time of
$\tau_{\rm var} \approx 2.2$~years, which is consistent with our main
observational result that the coronal line gas emission in NGC~4151
has varied in an observing period of $\sim 8-11$~years, but only
weakly so relative to the variability observed for the broad lines and
hot dust emission.

But how can we explain our other observational result that during half
the time period covered by the data the variable coronal lines showed
the opposite trend to the flux changes observed for the broad lines
and hot dust emission, i.e. their emission decreased instead of
increased? In order to answer this question, we first need to
understand how the coronal line region in NGC~4151 relates to the
BELR, the dusty torus and the low-ionisation NELR. For this purpose,
we have plotted in Fig. \ref{dens} the run of gas number density with
radius for a relationship $n \propto r^{-2}$, which keeps the
ionisation parameter constant. We have scaled this relationship to the
observed lag time for the hydrogen BELR (of about a week) and the
typical density for this emission region of $n_e \sim
10^{10}$~cm$^{-3}$. This scaling gives at the observed lag time for
the hot dust (of $\sim 50$~days) a density of $n_e \sim
10^8$~cm$^{-3}$, which is consistent with what is currently assumed
for the dusty torus material. From the relationship in
Fig. \ref{dens}, we see that, at the density estimated for the coronal
line region of $n_e \sim 10^3$~cm$^{-3}$, a ionisation parameter
typical of the BELR is reached at distances from the ionising source
of $\sim 55$~light~years, which is in the range observed for the
spatially resolved low-ionisation NELR. However, the coronal line
region in NGC~4151 appears to require a much higher ionisation
parameter than the hydrogen BELR for its lines to form, which, at a
given radius, can only be achieved by a considerable drop in
density. Therefore, it seems that the coronal line region is not part
of a continuous gas distribution but rather an independent entity.

One possibility proposed by \citet{Pier95} is that the coronal line
region is a layer on the inner part of the dusty torus that becomes an
efficient coronal line emitter only when evaporated in an X-ray heated
wind. If this scenario applies to NGC~4151, the wind will have
undergone adiabatic expansion from its launch location at the inner
face of the torus until the gas density and distance from the ionising
source are optimal to give the required high ionisation parameter. In
the process, the coronal line gas will have cooled. In
Fig. \ref{dusttemp}, we present tentative evidence for this
scenario. In the bottom panel, we compare the temperature of the
coronal line gas as measured by the line ratio
\FeVII~$\lambda$6087/\FeVII~$\lambda$3759 (the higher its value, the
lower the gas temperature; see Fig. \ref{FeVIItemp}) with that of the
hot dust. We find that the two temperatures behave in opposite ways;
the temperature of the coronal line gas is high when the hot dust
temperature is low and vice versa. Only the data from January 2007 is
an exception to this trend and might indicate a reddening event, which
could also explain the decreased flux of the H$\alpha$ broad line
relative to that of the Pa$\beta$ broad line for this epoch (see
Section \ref{iroptvar}). In the top panel of Fig. \ref{dusttemp}, we
show the continuum flux at rest-frame wavelength of $\sim 1$~$\mu$m,
which samples the accretion disc luminosity, versus the hot dust
temperature. A clear correlation is apparent, which indicates that the
change in temperature for the hot dust is due to direct heating by the
central ionising source. Therefore, the increased AGN radiation that
heats the dusty torus appears to increase the cooling efficiency of
the coronal line gas. In the scenario of \citet{Pier95}, the dusty
clouds will be evaporated in an X-ray heated wind more efficiently for
higher AGN luminosities, which will lead to an increase in mass
outflow rate but also to a stronger adiabatic expansion and so
cooling.

\section{Summary and conclusions}

We have presented the first extensive study of the coronal line
variability in an AGN. Our data set for the nearby, well-known source
NGC~4151 is unprecedented in that it includes six epochs of
quasi-simultaneous optical and near-IR spectroscopy spanning a period
of $\sim 8$~years and five epochs of X-ray spectroscopy overlapping in
time with it. Our main results are as follows.

\smallskip

(i) The variability behaviour observed for the broad emission lines
and hot dust emission was not mirrored in any of the coronal
lines. The coronal lines varied only weakly, if at all. Specifically,
in the first four years sampled by the data, the coronal lines either
did not vary significantly or showed the opposite behaviour to that of
the broad lines and hot dust emission, whereas after that only the
coronal lines with the highest ionisation potentials showed a
variability behaviour similar to that of the broad lines and hot dust
emission, but with a much reduced amplitude (a lower flux change per
year by a factor of $\sim 2-4$).

(ii) We have applied plasma diagnostics to the optical \FeVII~and
X-ray \OVII~emission lines in order to constrain the gas number
density, temperature and ionisation parameter of the coronal line
region. We find that this gas has a relatively low density of $n_e
\sim 10^3$~cm$^{-3}$ and requires a relatively high ionisation
parameter of $\log U \sim 1$.

(iii) We estimate the distance of the coronal line region in NGC~4151
from the central ionising source for the above gas density to be
$\tau_{\rm lt} \sim 2.0$~light years. This value, which is well below
the spatial resolution provided by current direct imaging results,
puts this region well beyond the hot inner face of the obscuring dusty
torus (of $\sim 2$~light months). Together with the recombination time
this results in a characteristic variability time scale of $\tau_{\rm
  var} \approx 2.2$~years, which is consistent with our main
observational result that the coronal line gas emission has varied in
an observing period of $\sim 8-11$~years, but only weakly so relative
to the variability observed for the broad emission lines and hot dust
emission.

(iv) Since the coronal line region requires a much higher ionisation
parameter than the BELR, it cannot be part of a continuous gas
distribution but is rather an independent entity. One possibility is
that this region is a layer on the inner part of the dusty torus that
becomes an efficient coronal line emitter only when evaporated in an
X-ray heated wind \citep{Pier95}. We present tentative evidence for
this scenario in the form of a temperature anti-correlation between
the coronal line gas and hot dust, which indicates that the increased
AGN radiation that heats the dusty torus appears to increase the
cooling efficiency of the coronal line gas, most likely due to a
stronger adiabatic expansion.

\smallskip

In a future paper, we plan to present a similar study of the coronal
line variability for the well-known AGN NGC~5548.

\section*{Acknowledgments}

We thank J\"org-Uwe Pott and Kirsten Schn\"ulle for making their IRTF
spectrum from 2010 available to us in electronic format. HL is
supported by a European Union COFUND/Durham Junior Research Fellowship
(under EU grant agreement number 267209). KCS thanks the astronomy
group at Durham University for its hospitality during a collaborative
visit supported by a Santander Mobility Grant. GJF acknowledges
support by NSF (1108928, 1109061, and 1412155), NASA (10-ATP10-0053,
10-ADAP10-0073, NNX12AH73G, and ATP13-0153), and STScI (HST-AR-13245,
GO-12560, HST-GO-12309, GO-13310.002-A, and HST-AR-13914), and to the
Leverhulme Trust for support via the award of a Visiting Professorship
at Queen's University Belfast (VP1-2012-025).

\bibliography{/Users/herminelandt/references}

\bsp
\label{lastpage}

\end{document}